\documentclass[twoside]{dis04}
\def\runauthor{M. Wing}
\def\shorttitle{Charm production in deep inelastic scattering}
\begin{document}

\title{CHARM PRODUCTION IN DEEP INELASTIC SCATTERING
}

\author{MATTHEW WING \\ (On behalf of the ZEUS Collaboration)}

\address{Bristol University, ZEUS, DESY\\
Notkestrasse 85, 22607 Hamburg, Germany\\
E-mail: wing@mail.desy.de }

\maketitle

\abstracts{Precise measurements of charm production in deep inelastic 
scattering are presented and compared with next-to-leading-order QCD. The data 
show sensitivity to the choice of parametrisation of the gluon density in 
the proton and could be used in future global fits of the parton densities.
}

\section{Introduction} 

Charm quarks are produced copiously in deep inelastic scattering (DIS) at HERA.
At sufficiently high photon virtualities, $Q^2$, the production of charm
quarks constitutes up to $30\%$ of the total cross
section~\cite{epj:c12:35,pl:b528:199}. Previous measurements of $D^*$ cross
sections~\cite{epj:c12:35,pl:b528:199,pl:b407:402,np:b545:21} indicate that the
production of charm quarks in DIS in the range $1 < Q^2 < 600$ GeV$^2$ is
consistent with QCD calculations in which charm is
produced through the boson-gluon-fusion mechanism. This implies that the
charm cross section is directly sensitive to the gluon density in the proton.

In this paper, measurements of the $D^*$ cross section~\cite{zeus_dstar_paper9800} 
are presented with improved precision and in a kinematic region extending to higher 
$Q^2$ than the previous ZEUS results~\cite{epj:c12:35}. The cross sections are 
compared to a next-to-leading-order (NLO) QCD calculation using various parton 
density functions (PDFs) in the proton. In particular, the data are compared to 
calculations using the recent ZEUS NLO fit~\cite{pr:d67:012007}, in which 
the parton densities in the proton are parametrised by performing fits to 
inclusive DIS measurements from ZEUS and fixed-target experiments. The 
cross-section measurements are used to extract the charm contribution, 
$F_2^{c\bar{c}}$, to the proton structure function, $F_2$.

\section{Results}

An integrated luminosity of 82~pb$^{-1}$ was used, more than double that of the 
previous ZEUS 
result~\cite{epj:c12:35}. Events were selected in a DIS regime defined 
by the kinematic region \mbox{$1.5 < Q^2 < 1000 \ {\rm GeV^2}$} and the 
inelasticity, $y$, restricted to \mbox{$0.02 < y < 0.7$}. 
Charm quarks were identified in these events by tagging a $D^*$ meson in the decay 
channel \mbox{$D^* \to D^0 \pi_s \to K\pi\pi_s$}. Transverse momentum, 
$p_T(D^*)$, and pseudorapidity, $\eta(D^*)$, requirements on the $D^*$ meson, 
governed mainly by the acceptance of the tracking detector, were 
\mbox{$1.5 < p_T(D^*) < 15$~GeV} and \mbox{$|\eta(D^*)| < 1.5$}. After all cuts, 
5545~$\pm$~129 $D^*$ mesons remained for cross-section measurements.

The $D^*$ production rate, $r = N/\mathcal{L}$, in the $e^-p$ 
data set is systematically higher than that in the $e^+p$ data set. This difference
increases with $Q^2$; for example, the  ratio of the rates, $r^{e^-p}/r^{e^+p}$,
is equal to $1.12 \pm0.06$ for \mbox{$1.5<Q^2<1000$ GeV$^2$}, while for
\mbox{$40<Q^2<1000$ GeV$^2$} it is $1.67\pm 0.21$ (only statistical errors are
given). Such a difference in production cross sections is not expected from known
physics processes. Many checks were performed to try and understand this observed 
difference; none gave an indication of its source. Therefore, the difference in 
rate is assumed to be a statistical fluctuation and the two sets of data were 
combined for the final results. An increased sample of $e^-p$ data can resolve 
the issue.

\begin{figure}[!thb]
\begin{center}
~\epsfig{file=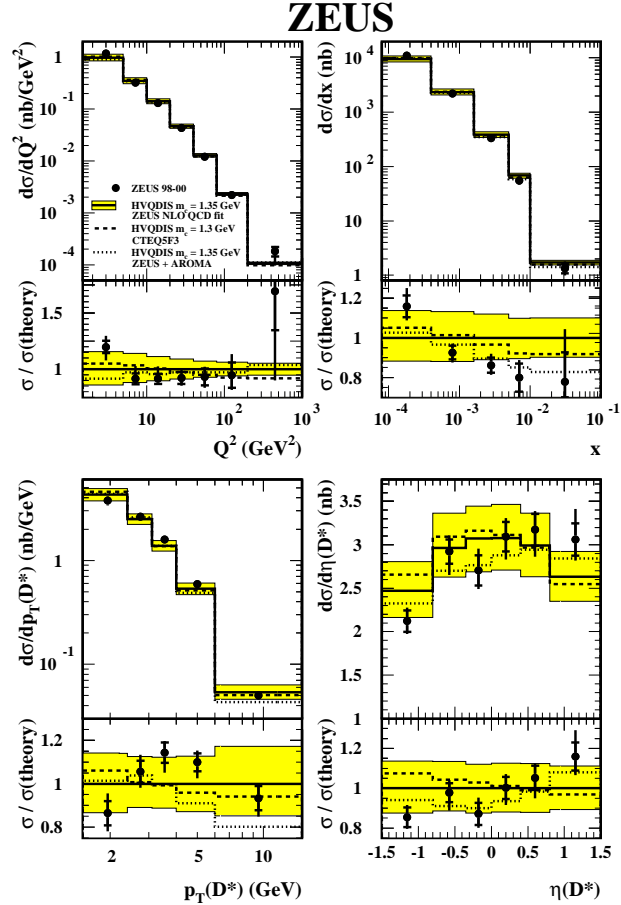,height=12cm}
\caption[*]{Differential cross sections compared to predictions from NLO QCD. 
The ratios of the cross 
sections to the central NLO QCD predictions are also shown beneath each plot.
\label{fig:xsections}}
\end{center}
\end{figure}

The differential $D^*$ cross sections as a function of $Q^2$, $x$, $p_T(D^*)$ and 
$\eta(D^*)$ are shown in Figure~\ref{fig:xsections} compared with predictions from 
an NLO QCD calculation~\cite{hvqdis}. Predictions are shown for two PDFs: ZEUS NLO 
and CTEQ5F3~\cite{cteq5}; and two hadronisation schemes: the Peterson function 
and the Lund string fragmentation~\cite{lund} as in {\sc Aroma}~\cite{aroma}. 
For all cross sections, the NLO predictions give a reasonable description of the 
data. The NLO calculation does, however, exhibit a somewhat different shape, 
particularly for $d\sigma/dx$, where the NLO is below the data at low $x$ and 
above the data at high $x$. The predictions using CTEQ5F3 instead of the ZEUS NLO 
fit, or using {\sc Aroma} for the hadronisation instead of the Peterson function, 
give better agreement with the data. The 
prediction using the ZEUS NLO fit gives a better description than that using 
CTEQ5F3 (and also better than the prediction using GRV98-HO~\cite{grv98}, not shown) 
for the cross-section $d\sigma/d\eta(D^*)$. A better description is also achieved 
by using {\sc Aroma} for the hadronisation, 
although, in this case, $d\sigma/dp_T(D^*)$ is not so well described. Further 
refinement of NLO QCD fits and even the use of these data in future fits may 
achieve a better description.

The open-charm contribution, $F_2^{c \bar{c}}$, to the proton structure-function 
$F_2$ can be defined in terms of the inclusive double-differential $c\bar{c}$ cross
section in $x$ and $Q^2$ by
\begin{equation}
\frac{d^2\sigma^{c\bar{c}} (x, Q^2)}{dxdQ^2} =
\frac{2\pi\alpha^2}{x Q^4}
\{ [1+(1-y)^2] F_2^{c\bar{c}}(x, Q^2) - y^2 F_L^{c\bar{c}}(x, Q^2) \} .
\label{eq:nc_charm}
\end{equation}
In this analysis, the $c\bar{c}$ cross section is obtained by measuring the $D^*$
production cross section and employing the hadronisation fraction
$f(c \rightarrow D^\ast)$ to derive the total charm cross section. Since only
a limited kinematic region is accessible for the measurement of $D^*$ mesons,
a prescription for extrapolating to the full kinematic phase space is needed. 
The measured $F_2^{c \bar{c}}$ in a bin $i$ is given by
\begin{equation}
F_{2,\rm meas}^{c\bar{c}}(x_i, Q^2_i) = \frac{\sigma_{i,\rm meas}(ep \rightarrow D^* X)}
                                      {\sigma_{i,\rm theo}(ep \rightarrow D^* X)}
                                      F_{2,\rm theo}^{c\bar{c}}(x_i, Q^2_i),
\end{equation}
where $\sigma_i$ are the cross sections in bin $i$ in the measured region of
$p_T(D^*)$ and $\eta(D^*)$. The value of $F_{2, \rm theo}^{c \bar{c}}$ was 
calculated from the NLO coefficient functions~\cite{pr:d67:012007}. In this 
calculation, the same parton densities, charm mass ($m_c =$ 1.35 GeV), and 
factorisation and renormalisation scales ($\sqrt{4m_c^2 + Q^2}$) have been 
used as for the NLO QCD calculation of the differential cross sections. The 
hadronisation was performed using the Peterson fragmentation function. Typical 
extrapolation factors to the full phase space in $p_T(D^*)$ and $\eta(D^*)$ 
are 4.7 at low $Q^2$ and 1.5 at high $Q^2$.

\begin{figure}[!thb]
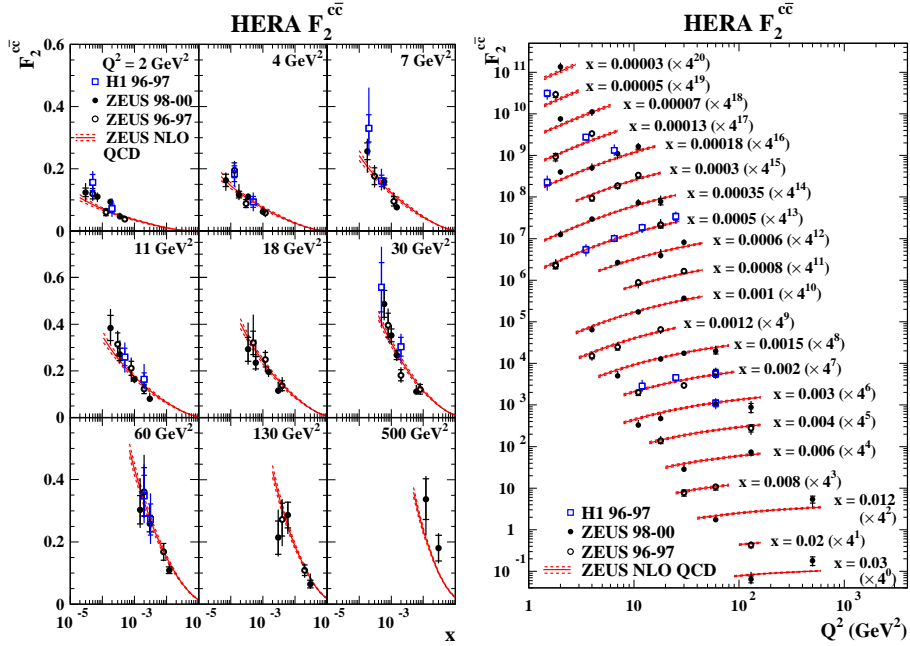

\begin{center}
~\epsfig{file=fig6_h1_zeus.epsi,height=8.5cm}
~\epsfig{file=fig7_h1_zeus.epsi,height=8.5cm}
\caption[*]{The measured $F_2^{c \bar{c}}$ as a function of $x$ for fixed $Q^2$ 
and as a function of $Q^2$ for fixed $x$. The current data are compared with 
previous H1 and ZEUS measurements the predictions from the ZEUS NLO QCD fit.
\label{fig:f2charm}}
\end{center}
\end{figure}

The data for $F_2^{c \bar{c}}$ as a function of $x$ for fixed $Q^2$ are 
compared in Figure~\ref{fig:f2charm} with previous measurements from 
H1~\cite{pl:b528:199} and ZEUS~\cite{epj:c12:35} and with the ZEUS NLO fit. 
The three sets of data are consistent. The values of $F_2^{c \bar{c}}$ are also 
presented as a function of $Q^2$ at fixed values of $x$ and compared with the 
ZEUS NLO fit in Figure~\ref{fig:f2charm}. The data rise with increasing $Q^2$, 
with the rise becoming steeper at lower $x$, demonstrating the property of 
scaling violation in charm production. The prediction describes the data well
for all $Q^2$ and $x$. The uncertainty on the theoretical prediction is that
from the PDF fit propagated from the experimental uncertainties of the fitted
data. At the lowest $Q^2$, the uncertainty in the data is comparable to the
PDF uncertainty shown. This implies that the double-differential cross sections 
could be used as an additional constraint on the gluon density in the proton.

\section{Conclusions}

The production of $D^*$ mesons has been measured in DIS at
HERA. Predictions from NLO QCD are in reasonable agreement with the measured 
cross sections, which show sensitivity to the choice of PDF and hence the gluon 
distribution in the proton. The ZEUS NLO PDF, which was fit to recent inclusive 
DIS data, gives the best description of the $D^*$ data. In particular, this is 
seen in the cross-section $d\sigma/d\eta(D^*)$. The double-differential cross 
section in $y$ and $Q^2$ has been measured and used to extract the open-charm 
contribution to $F_2$. Since, at low $Q^2$, the 
uncertainties of the data are comparable to those from the PDF fit, the measured 
differential cross sections should be used in future fits to 
constrain the gluon density.

\end{document}